 \documentclass[preprint,review,12pt]{elsarticle}

 \usepackage{graphics}
 \usepackage{graphicx}
 \usepackage{epsfig}

 \usepackage{amssymb}
 \usepackage{amsthm}
 \usepackage{float}
 \usepackage[nodots]{numcompress}

\journal{Ecological Complexity and Accepted for publication (in 2013)}

\begin{document}

\begin{frontmatter}

\author{C.~Jeevarathinam \fnref{label1}}
\ead{c.jeeva1987@gmail.com}
\author{S.~Rajasekar \fnref{label1}}
\ead{rajasekar@cnld.bdu.ac.in}
\author{Miguel A.F.~Sanju\'{a}n \fnref{label2}}
\ead{miguel.sanjuan@urjc.es}

\title{Vibrational resonance in groundwater-dependent plant ecosystems}

\address[label1]{School of Physics, Bharathidasan University, 
Tiruchirappalli  620 024, Tamilnadu, India}
\address[label2]{Nonlinear Dynamics, Chaos and Complex Systems Group, Departamento de F\'{i}sica, \\ Universidad Rey Juan Carlos, Tulip\'{a}n s/n, 28933 M\'{o}stoles, Madrid, Spain}

\begin{abstract}
We report the phenomenon of vibrational resonance in a single species and a two species models of groundwater-dependent plant ecosystems with a biharmonic oscillation (with two widely different frequencies $\omega$ and $\Omega$,  $\Omega \gg \omega$) of the water table depth. In these two systems, the response amplitude of the species biomass shows multiple resonances with different mechanisms. The resonance occurs at both low- and high-frequencies of the biharmonic force. In the single species bistable system, the resonance occurs at discrete values of the amplitude $g$ of the high-frequency component of the water table. Furthermore, the best synchronization of biomass and its carrying capacity with the biharmonic force occurs at the resonance. In the two species excitable and time-delay model, the response amplitude $(Q)$ profile shows several plateau regions of resonance, where the period of evolution of the species biomass remains the same and the value of $Q$ is inversely proportional to it. The response amplitude is highly sensitive to the time-delay parameter $\tau$ and shows two distinct sequences of resonance intervals with a decreasing amplitude with $\tau$.   

\end{abstract}


\end{frontmatter}

\section{Introduction}
\label{1}
The study of the response of a system to a small variation of environmental changes is important since environmental drivers often fluctuate. The fluctuation can be periodic or nonperiodic (noise). In certain ecosystems, a small change in one or more environmental parameters leads to considerable changes on their structure and function. It has been noted that many systems have relatively high levels of diversity for an intermediate level of a disturbance (Roxburgh et al., 2004). The impact of the environmental variations/fluctuations has been analysed on food web stability (Vasseur and Fox, 2007), species coexistence in Holt-McPeek systems (Lai and Liu, 2005) and the stability of recovery (Steneck et al., 2002). Without invoking interaction  between environmental noise and competition, it has been shown that environmental fluctuations enhance coexistence of species which either prefer or tolerate different environmental conditions (D'Odorico et al., 2008).

In the present work, we consider groundwater-dependent plant ecosystems. It is important to analyse the influence of the variation of various environmental factors, particularly, the changes in the water table depth, in order to get a deep understanding of the response of various ecosystems. We point out that vegetation-water table interactions is very common in many ecosystems like wetlands, salt marshes and riparian forests. As a matter of fact, it is considered as one of the key mechanisms influencing the dynamics of vegetation (Naumburg et al., 2005; Elmore et al., 2006; Munoz-Reinoso and de Castro, 2005). Appropriate theoretical models are of  great use for exploring various possible dynamics that can emerge from vegetation-water table interactions. In this connection, (Ridolfi et al., 2006, 2007) have proposed two vegetation-water table models based on realistic ecological assumptions. The first model describes the vegetation biomass dynamics of only one species (dominant species). In this model the rate of change of the species biomass depends on the existing biomass and the carrying capacity of the system. The resultant model is a first-order nonlinear ordinary differential equation with a periodic driver. It accounts for multistable states in the dynamics of wetland forests and riparian ecosystems (Scheffer et al., 2001). The second model describes the two phreatophyte species interacting with a water table. In these two models, one of the factors that can change the carrying capacity of biomass is the depth of the water table. Change in the water depth due to seasonal rainfall oscillations and other sources is represented by a periodic function of time. It is also found to display coexistence of two species and chaotic dynamics (Ridolfi et al., 2007).

The influence of the environmental variability, treated as a disturbance or a kind of noise, in the above two models has been analyzed recently by Borgogno and his co-workers (Borgogno et al., 2012). Specifically, they have shown the occurrence of stochastic and coherence resonances. When a bistable or an excitable system driven by a weak periodic force is subjected to an additive noise, it can exhibit an enhanced response at an optimal noise intensity. This phenomenon is termed as stochastic resonance (Gammaitoni et al., 1998; McDonnell et al., 2008).  Very recently, noise-induced spatio-temporal patterns in wetland vegetation dynamics have been reported (Scarsoglio et al., 2012).  In a subthreshold excitable system, a noise-induced resonance can be realized in the absence of external periodic driving and is known as coherence resonance (Pikovsky and Kurths, 1997). Interestingly, it has been shown that deterministic resonances can be observed even in monostable nonlinear systems driven by a biharmonic force in the absence of external noise and is called vibrational resonance 
(Landa and McClintock, 2000). The analysis of vibrational resonance has received a great deal of attention in recent years.  Particularly, its occurrence has been investigated in a spatially extended system in the presence of noise  (Zaikin et al., 2002), Duffing oscillator (Blekhman and Landa, 2004), two-coupled overdamped anharmonic oscillators (Gandhimathi et al., 2006) and monostable systems (Jeyakumari et al., 2009).  Experimental evidence of vibrational resonance was demonstrated in analog simulations of the overdamped Duffing oscillator (Baltanas et al., 2003),  a bistable optical cavity laser (Chizhevsky and Giacomelli, 2006) and an excitable electronic circuit with Chua's diode (Ullner et al., 2003).  The influence of time-delayed feedback on vibrational resonance was studied numerically (Yang and  Liu, 2010) and theoretically (Jeevarathinam et al., 2011).  Further, biharmonic force induced enhanced signal propagation was found to occur in one-way coupled systems (Yao and  Zhan, 2010) and in a coupled network of excitable neuronal systems (Yu et al., 2011).

In the present work, we consider the two groundwater-dependent plant ecosystem models of (Ridolfi et al., 2006, 2007) and investigate the emergence of vibrational resonance. It has been pointed out that the dynamics of vegetation have a time scale greater than one season and much greater than man-induced periodic disturbances (Ridolfi et al., 2007). We wish to mention that high-frequency oscillation of beach water table due to wave runup and rundown has been observed and analysed (Waddell, 1976; Li et al., 1997).  Interestingly, similar high-frequency oscillation of underground water table (in addition to the low-frequency periodic oscillation of water table due to seasonal variation)  can occur due to evaporation, inflow  and outflow of water and temperature fluctuation.  It can also be artificially realized through irrigation or pumping from the aquifer.  Furthermore, a water table rise and drop can be induced by vegetation removal and planting respectively. Thus, planting additionally short-lived species interacting weakly with species $A$ can also lead to a high-frequency variation of the water depth.  Therefore, it is realistic to include a biharmonic force in the water table with two well-separated frequencies. The first model describing the dynamics of the biomass $V$ of a single species has bistable states. When the biharmonic force is included in the water depth the system shows an oscillatory variation of $V$. As the amplitude of the high-frequency force is varied, the system exhibits multiple vibrational resonance with a decreasing response amplitude at successive resonances for certain range of fixed values of amplitude of low-frequency force. The second model describes the dynamics of two species, say $A$ and $B$, interacting with the water table, and where the evolution of $B$ depends on $A(t-\tau)$ and $\tau$ is the time-delay parameter. Unlike the single species model, the two species model is an excitable system (such a system  have only one stable equilibrium state, but external perturbations above a certain threshold can induce large excursions in phase space, which takes the form of spikes or pulses).  For a fixed time-delay, both $A$ and $B$ display a certain number of resonances when the amplitude of the high-frequency force is varied. The resonance profiles of $A$ and $B$ are similar except that at resonance the amplitude of $A$ is always much higher than that of $B$. Here the resonance intervals are not sharp but wide. The response amplitude is inversely proportional to the period of the variation of $A$ and $B$. The delay parameter $\tau$ has a strong influence on the response amplitude. The response amplitude at successive alternative resonances decreases when the value of the delay parameter increases.

\section{Vibrational resonance in a single species model}
To start with, first we briefly introduce the model (Ridolfi et al., 2006, 2007) in order to prepare the readers for the study of vibrational resonance.

\subsection{Description of the model}
The dynamics of phreatophyte biomass $V$ of a single species (or total plant biomass neglecting interspecies interactions) is expressed as (Ridolfi et al., 2006, 2007)
\begin{equation} 
 \label{eq1}
   \frac{ {\mathrm{d}}V }{ {\mathrm{d}}t }
     = V \left( V_{{\mathrm{cc}}} - V \right) ,
\end{equation}
where the growth rate of $V$ is assumed to be proportional to the existing biomass and the available resources $V_{{\mathrm{cc}}}-V$ with $V_{{\mathrm{cc}}}$ being the carrying capacity of the ecosystem, that is, the maximum amount of vegetation sustainable with the available resources. Based on experimental evidences, an appropriate form of $V_{{\mathrm{cc}}}$ shows a quadratic dependence on the water table depth $d$. Taking into the effect of periodic oscillations in the rainfall regions leading to periodic variations of water table depth, (Borgogno et al., 2012) considered the form of $V_{{\mathrm{cc}}}$ as
\begin{equation}  
 \label{eq2}
   V_{{\mathrm{cc}}} 
     = \left\{ \begin{array}{ll}
        a \left[ d(t) - d_{{\mathrm{inf}}} \right] 
        \left[ d_{{\mathrm{sup}}} - d(t) \right],
         & {\mathrm{ if}} \; d_{{\mathrm{inf}}}
                 < d < d_{{\mathrm{sup}}}  \\
              0, & {\mathrm{otherwise}}.  \end{array} \right.
\end{equation}
The form of $V_{{\mathrm{cc}}}$ given by Eq.~(\ref{eq2}) corresponds to the case of phreatophyte vegetation that depends on water uptake from the groundwater.  In Eq.~(\ref{eq2}) $d(t)$ is the water table depth, $a$ is the sensitivity of carrying capacity to changes in the water table depth, $d_{{\mathrm{inf}}}$ is the threshold of vegetation tolerance to shallow water tables and insufficient aeration of the root zone and $d_{{\mathrm{sup}}}$ is the threshold of water depth below which tap-roots cannot extract water.  The water table depth is given by
\begin{equation}
 \label{eq3}
    d(t) = d_0 + \beta V + F(t),
\end{equation}
where $d_0$ is the water depth in the absence of vegetation, $\beta$ is the sensitivity of the water table to the presence of vegetation and $F(t)$ describes the oscillatory variation of the water table. The choice $F(t)=f \cos\omega t$ is considered in (Borgogno et al., 2012). In the present work we choose $F(t)$ as a biharmonic force with two widely differing frequencies:
\begin{equation}
 \label{eq4}
    F(t) = f \cos \omega t + g \cos \Omega t, 
               \quad \Omega \gg \omega.
\end{equation}
The explicit time-dependent variation of the water depth can be natural due to seasonal rainfall oscillations or man-induced perturbations (pumping from an aquifer). 

\begin{figure}[t]     
\begin{center}
\includegraphics[width=0.9\linewidth]{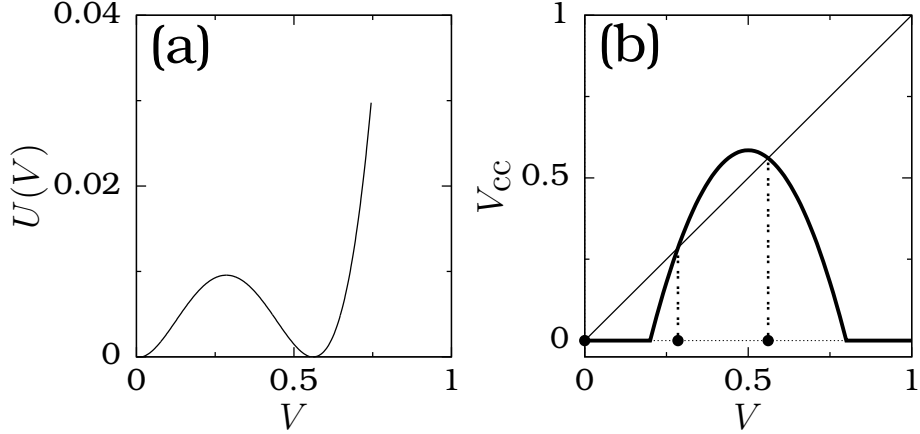}
\end{center}
\caption{(a) The potential $U(V)$ of the system (\ref{eq1}) for $f=0$, $g=0$. The values of the parameters are $d_0=0.5{\mathrm{m}}$, $\beta=0.5{\mathrm{m}}$, $a=26{\mathrm{m}}^{-1}$, $d_{{\mathrm{inf}}}=0.6{\mathrm{m}}$ and $d_{{\mathrm{sup}}}=0.9{\mathrm{m}}$  the values used in (Borgogno et al., 2012). (b) Variation of $V_{{\mathrm{cc}}}$ with $V$ (thick curve). The equilibrium points are the intersections of the bisector (thin straight-line) with the $V_{{\mathrm{cc}}}$ curve. The locations of the equilibrium points are marked by solid circles.}
\label{fig1}
\end{figure}

The potential $U(V)$ defined through ${\mathrm{d}} V / {\mathrm{d}}t = -{\mathrm{d}} U / {\mathrm{d}}V$ in the absence of $F(t)$ is depicted in Fig.~\ref{fig1}a where $d_0=0.5{\mathrm{m}}$, $\beta=0.5{\mathrm{m}}$, $a=26{\mathrm{m}}^{-1}$, $d_{{\mathrm{inf}}}=0.6\mathrm{m}$ and $d_{{\mathrm{sup}}}=0.9{\mathrm{m}}$ the values of the parameters used in (Borgogno et al., 2012).  $U(V)$ is of a double-well form. The equilibrium states can be obtained by setting  ${\mathrm{d}} V / {\mathrm{d}} t=0$. $V^*_0=0$, representing the unvegetated state, is an equilibrium point. The other equilibrium states correspond to $V_{{\mathrm{cc}}}=V$. In the plot between $V_{{\mathrm{cc}}}$ versus $V$ the intersection points of the line $V_{{\mathrm{cc}}}=V$ with the curve of  $V_{{\mathrm{cc}}}$ are the equilibrium states. Figure~\ref{fig1}b shows  $V_{{\mathrm{cc}}}$ versus $V$. The equilibrium states are  $V^*_0=0$,  $V^*_{{\mathrm{u}}}=0.28526$ and  $V^*_{{\mathrm{s}}} = 0.56090$. $V^*_{{\mathrm{u}}}$ is the local maximum of $U(V)$ and is an unstable state.  $V^*_0$ and  $V^*_{{\mathrm{s}}}$ are two local minima of $U(V)$ and are stable states. When the periodic function $F(t)$ is taken into consideration, then the effect of $F(t)$ is to periodically modulate the potential $U(V)$.

\subsection{Multiple resonance induced by the biharmonic force} 
In the $F(t)$ given by Eq.~(\ref{eq4}), we have assumed that $\Omega \gg \omega$. In this case due to the difference in time scales of the low-frequency oscillation $f \cos \omega t$ and the high-frequency oscillation $g \cos \Omega t$, the solution of the system (\ref{eq1}) consists of a slow variation of $V(t)$ denoted by $V_{{\mathrm{slow}}}(t)$ and a fast variation $V_{{\mathrm{fast}}}(t,\Omega t)$.   We denote $Q_{\omega}$ and $Q_{\Omega}$ as the response amplitudes of $V$ at the frequencies $\omega$ and $\Omega$ respectively. A theoretical approach has been developed to obtain an analytical expression for $Q$ for certain class of oscillators (Landa and McClintock, 2000; Blekhman and Landa, 2004). The threshold existing between the carrying capacity $V_{{\mathrm{cc}}}$ and $d$ in Eq.~(\ref{eq2}) makes it difficult to use the theoretical approach to investigate the vibrational resonance. Therefore, we compute both $Q_{\omega}$ and $Q_{\Omega}$ from the numerical solution of the Eq.~(\ref{eq1}). From $V(t)$ the sine and cosine components $Q_{\omega,{{\mathrm{s}}}}$ and $Q_{\omega,{{\mathrm{c}}}}$ are computed from the equations
\begin{eqnarray}
 \label{eq5}
    Q_{\omega,{{\mathrm{s}}} }
       & = &  \frac{2}{nT} \int_0^{nT} V(t) \sin \omega t
             \, {\mathrm{d}} t , \\
 \label{eq6}
    Q_{\omega,{{\mathrm{c}}} }
       & = & \frac{2}{nT} \int_0^{nT} V(t) \cos \omega t
             \, {\mathrm{d}}t,
\end{eqnarray}
where $T = 2 \pi / \omega$ and $n$ is taken as, say, $200$. Then 
\begin{equation}
 \label{eq7}
   Q_{\omega} = \sqrt{ Q^2_{\omega,{\mathrm{s}}} 
         + Q^2_{\omega,{\mathrm{c}}} } {\Big{/}} f.
\end{equation}
Similarly, we can compute $Q_{\Omega}$ from the numerical solution $V(t)$.

\begin{figure}[!ht]     
\begin{center}
\includegraphics[width=0.52\linewidth]{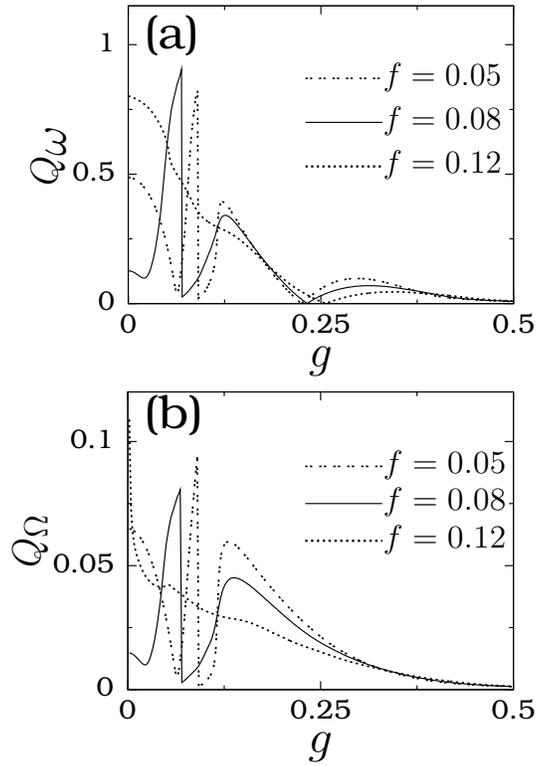}
\end{center}
\caption{Response amplitudes (a) $Q_{\omega}$ and (b) $Q_{\Omega}$ of plant biomass $V$ for three fixed values of the parameter $f$ with $\omega=1$ and $\Omega=10$.  The values of the other parameters in the system (\ref{eq1}) are as in Fig.~{\ref{fig1}}. }
\label{fig2}
\end{figure}

The value of  $f$ is significant  in observing vibrational resonance in the system (\ref{eq1}).  For a fixed value of $\omega$ and $g=0$, small amplitude oscillations  of $V$ occur about the coexisting stable equilibrium points $V_0^*$ and $V_{{\mathrm{s}}}^*$ for $\vert f \vert \ll 1$.  Then for fixed values of $f$ with $\vert f \vert \ll 1$ and $\Omega \gg \omega$, when $g$ is varied  both $Q_{\omega}$ and $Q_{\Omega}$ display one or more resonances depending upon the values of the parameters of the system.  For   $g>0.5$, both  $Q_{\omega}$ and $Q_{\Omega} \approx 0$ and there is no further resonance.  Therefore, we present the results for $0<g<0.5$.  Figure~{\ref{fig2}} shows the dependence of  $Q_{\omega}$ and $Q_{\Omega}$ for three fixed values of $f$ with  $\omega=1$ and $\Omega=10$.   
In all the cases a nonmonotonic variation of both $Q_{\omega}$ and $Q_{\Omega}$ occurs.  In Fig.~{\ref{fig2}} $Q_{\omega}$ and $Q_{\Omega}$ are $\approx 0$ for $g>0.5$.   For $f=0.08$ as $g$ increases from a small value,  $Q_{\omega}$ initially decreases then increases sharply, reaches a maximum at $g={g_{_{\mathrm{VR1}}}}=0.069$ and then sharply decreases to a lower value. As $g$ increases further,  $Q_{\omega}$ becomes maximum at two other values of $g$, namely $g={g_{_{\mathrm{VR2}}}}=0.126$ and $g={g_{_{\mathrm{VR3}}}}=0.312$. There are three resonances.  We note that a resonance at the low-frequency of the water table oscillation is induced by an appropriate  value of the amplitude $g$ of the high-frequency oscillation of the water table.  For this reason the above resonance phenomenon, that is the occurrence of a maximum of $Q_{\omega}$, is termed as {\emph{vibrational resonance}}.  
For $f=0.05$ also three resonances occur. For both $f=0.05$ and $0.08$ the first resonance is the dominant and the value of $Q$ at the successive resonances decreases. On the other hand, the width of the successive bell-shaped resonance curves becomes wider and wider. When the value of $f$ increases the resonance peaks move towards lower values of $g$. For $f$ values above a critical value, the first two resonance peaks disappear. Moreover, the value of $Q$ at $g=0$ is substantially enhanced with an increase in the values of $f$ producing a resonance without tuning.

In the nonlinear oscillators driven additively by a biharmonic force, $Q_{\Omega}$ monotonically increases when $g$ increases and there is no resonance-like variation of it. We wish to remark that in the system (\ref{eq1}), the biharmonic force is not an additive force but it is in the expression for $V_{{\mathrm{cc}}}$.  Therefore, one wish to know the response of the system (\ref{eq1}) and its relative strength at the high-frequency $\Omega$ of the water table oscillation.  For this purpose, we computed and presented the variation of $Q_{\Omega}$ with $g$.  Interestingly,  $Q_{\Omega}$ also displays  resonance. This is shown in Fig.~\ref{fig2}b.  The dependence of $Q_{\Omega}$ on $g$ is similar to that of $Q_{\omega}$.   For $f=0.05$ and $0.08$ there are two resonances with $Q_{\Omega}$ and both occur at the same values of $g$ at which $Q_{\omega}$ becomes maximum.  $Q_{\Omega}$ also shows resonance without tuning.  Though the resonances associated with the two frequencies $\omega$ and $\Omega$ occur at the same value of the control parameter $g$, at resonance $Q_{\omega}$ is much higher than $Q_{\Omega}$.  We note that $Q_{\omega}$ and $Q_{\Omega}$ are proportional to the Fourier coefficients of the periodic terms with the frequencies $\omega$ and $\Omega$ respectively in the Fourier series of $V(t)$.  Since the Fourier coefficients decay with increase in the frequency and because $\Omega \gg \omega$  the response amplitude $Q_{\Omega}$ is much less than $Q_{\omega}$ for each value of $g$ in Fig.~{\ref{fig2}}.  It is noteworthy to mention that at $g=g_{_{\mathrm{VR}}}$ an enhanced vegetation can be realized over two time intervals, one with the low-frequency $\omega$ and another with the high-frequency $\Omega$. These two frequencies need not be commensurable. 

\subsection{Dynamics of  $V$ and $V_{{\mathrm{cc}}}$ at and far from resonances}  
\begin{figure}[!ht]     
\begin{center}
\includegraphics[width=0.73\linewidth]{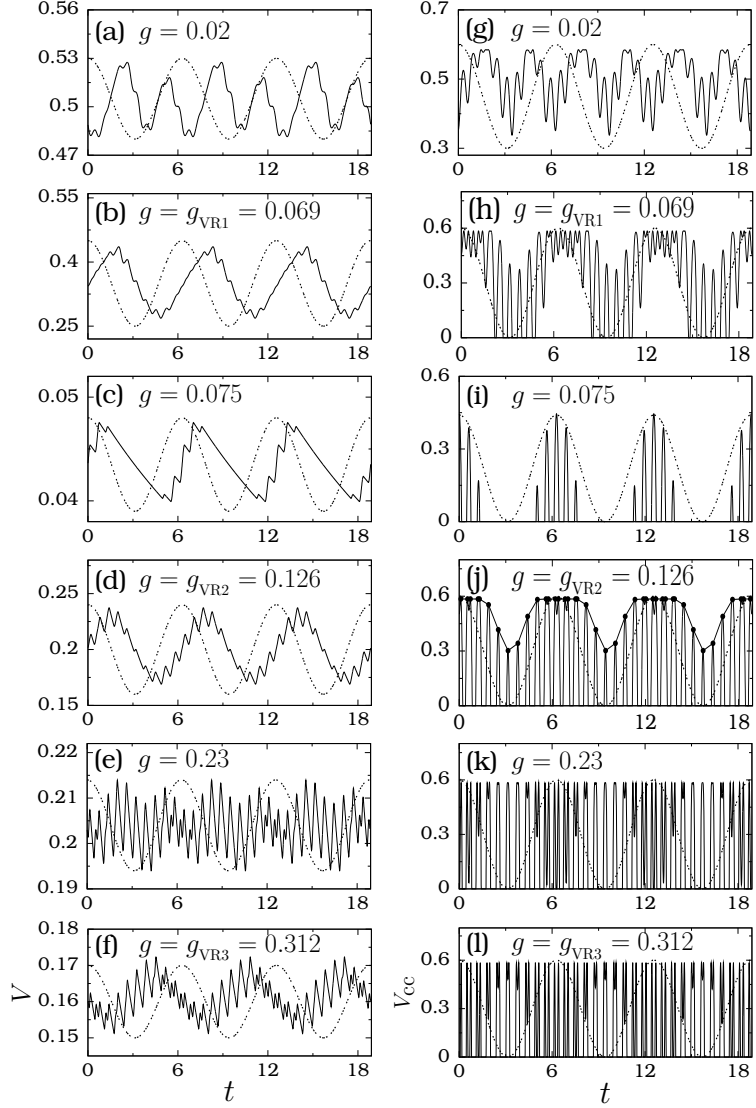}
\end{center}
\caption{The time evolution of (a)-(f) $V(t)$ and (g)-(l)  $V_{{\mathrm{cc}}}(t)$ of the system (\ref{eq1}) at few values of $g$ with $f=0.08$, $\omega=1$ and $\Omega=10$. The values of the other parameters are as in Fig.~{\ref{fig1}}.  The dashed curves represent $f \cos \omega t$. For comparative purpose the amplitude of $f \cos \omega t$ is suitably multiplied by a factor.}
\label{fig3}
\end{figure}

The occurrence  of resonance in the system (\ref{eq1}) can be understood by analyzing the influence of the control parameter $g$ on the evolution of the plant biomass $V$ and the carrying capacity $V_{{\mathrm{cc}}}$ of the ecosystem.  Figure {\ref{fig3}}  shows $V(t)$ and the carrying capacity $V_{{\mathrm{cc}}}(t)$ of vegetation biomass for several fixed values of $g$.   
In the absence of $F(t)$ there are two stable dynamical states $V^*_0$ and $V^*_{{\mathrm{s}}}$. The values of $f$ and $\omega$ are chosen in such a way that in the absence of a high-frequency oscillation of $F(t)$ the variable $V(t)$ oscillates either about $V^*_0$ or $V^*_{{\mathrm{s}}}$ depending upon the initial condition and there is no transition between these two states. When $g$ is varied from $0$ then for very small values two oscillatory states coexist. The equilibrium states about which oscillation occurs are perturbed by $F(t)$. For the entire range of values of $g$ considered in Fig.~{\ref{fig2}} the evolution of $V$ is periodic.   $V(t)$ is said to be periodic with period $t'$ if  $V(t+t')=V(t)$ for some finite and nonzero value of $t'$ after leaving initial transient evolution of it, say for example leaving $V(t)$ for $0 < t <100 \times T(=2 \pi/\omega)$.  The period of $V(t)$ is found to be  $T = 2 \pi / \omega$, the period of $F(t)$ given by Eq.~(\ref{eq4}).  In Fig.~{\ref{fig3}}a for $g=0.02$ far before the first resonance, for the initial condition chosen near $V^*_{{\mathrm{s}}}$, though both $V(t)$ and $f \cos \omega t$ are periodic with the same period $T$ the forms of both of them are different. Over one drive cycle both $V(t)$ and  $V_{{\mathrm{cc}}}(t)$ (see Fig.~{\ref{fig3}}g) have more than one dominant maximum while $F(t)$ has only one. There is no synchronization between $V$ (as well as $V_{{\mathrm{cc}}}$) and $f \cos \omega t$. For  $g={g_{_{\mathrm{VR1}}}}=0.069$ in Figs.~{\ref{fig3}}b and h the center of oscillation is shifted. $V$ and $V_{{\mathrm{cc}}}$ are synchronized with $f \cos \omega t$ (except that there is a phase difference). This feature of $V$ leads to the first resonance. The numerical analysis shows an absence of resonance at $g=0.069$ for the vegetation dynamics about the other low value stable state $(V^*_0)$. The value of $Q$ of the corresponding dynamics is very small.

Next, at $g=0.07$ the high-frequency oscillation of the water table induces a transition of $V$ about  $V^*_{{\mathrm{s}}}$ to $V^*_0$. When the vegetation biomass is close to $V^*_0=0$, it is easy to note from Eqs.~(\ref{eq2}) and (\ref{eq3}) that, $V_{{\mathrm{cc}}}=0$ for most of the time over a drive cycle of $F(t)$. Consequently, the dynamics is confined near  $V^*_0$. There is no transition between the two states $V^*_0$ and $V^*_{{\mathrm{s}}}$. This is shown in Figs.~{\ref{fig3}}c and i for $g=0.075$.  Though both $V$ and $V_{{\mathrm{cc}}}$  are quite synchronized with $f \cos \omega t$, there is no resonance because $V$ is trapped into the unvegetated state.  Because the center of oscillation of $V$ is shifted  from the vegetated state to the unvegetated state both $Q_{\omega}$ and $Q_{\Omega}$ make a sudden jump from a higher value to a smaller value at $g=0.07$ (see Fig.~{\ref{fig2}}).  When $g$ increases further, the time intervals in which $V_{\mathrm{cc}}=0$ decreases. In Fig.~{\ref{fig3}}j  corresponding to $g=0.126$ the total time over which  $V_{{\mathrm{cc}}}=0$ is $\approx T/2$ and the line joining the maxima of $V_{{\mathrm{cc}}}$ (indicated by solid circles connected by a line) varies sinusoidally and one can clearly notice synchronization between $V_{{\mathrm{cc}}}$ and $f \cos \omega t$. A second resonance occurs at this value of $g$. In Fig.~{\ref{fig3}}d $V$ oscillates about the perturbed equilibrium state $V^*_{{\mathrm{u}}}$. For $g=0.23$ rapid oscillations of $V$ and $V_{{\mathrm{cc}}}$ (Figs.~{\ref{fig3}}e and  k) take place and $Q_{\omega} \approx 0$. At $g=0.312$ a third resonance with $Q_{\omega}$ much less than that of the first two resonances occurs. However, we can clearly see the synchronization between $V$ and $f \cos \omega t$ in (Fig.~{\ref{fig3}}f). 

\begin{figure}[!ht]     
\begin{center}
\includegraphics[width=0.9\linewidth]{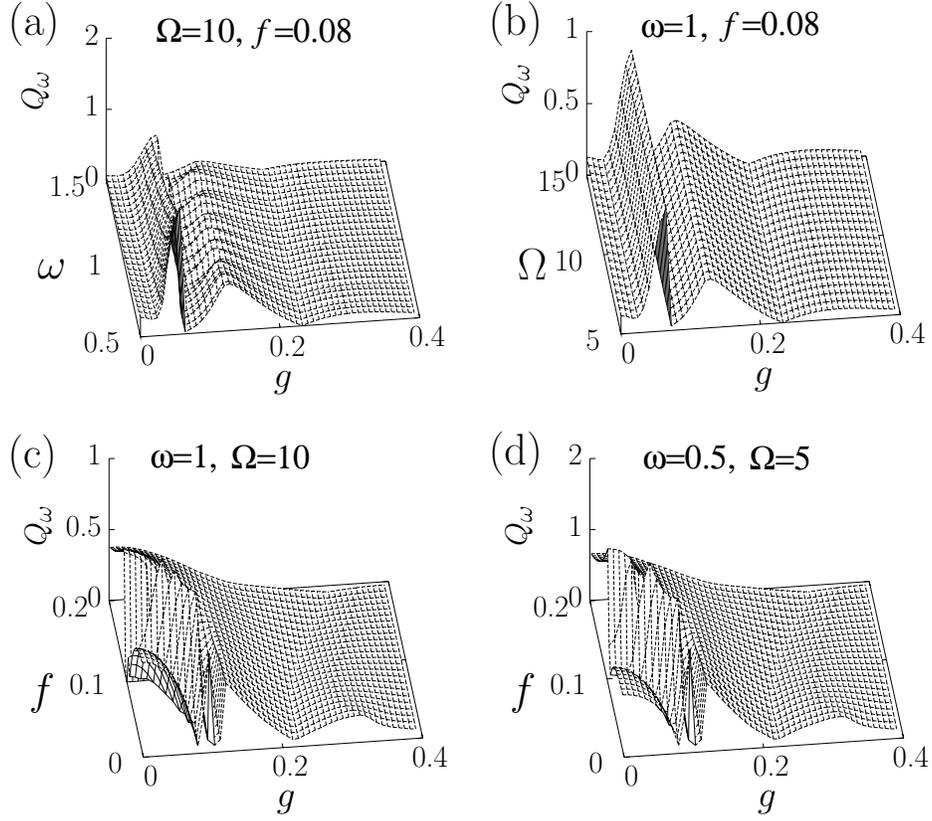}
\end{center}
\caption{Variation of $Q_{\omega}$ as a function of (a) $\omega$ and $g$ for $\Omega=10$, $f=0.08$, (b) $\Omega$ and $g$ for $\omega=1$ and $f=0.08$,  (c)-(d) $f$ and $g$ for two sets of fixed values of $\omega$ and $\Omega$.    The values of the other parameters are as in Fig.~{\ref{fig1}}.  }
\label{fig4}
\end{figure}

In this section so far we reported the results for $\omega=1$, $\Omega=10$ and $f=0.08$.  We numerically studied the occurrence of resonance for a wide range of fixed values  of $\omega$, $\Omega$ and $f$ and thereby varying the control parameter $g$.  Figures {\ref{fig4}}a and b show the variation of $Q_{\omega}$ with $g$ for $\omega \in [0.5,\,1.5]$ where $\Omega=10$ and $\Omega \in [5,\,15]$ where $\omega=1$ respectively.   Three resonances occur for a wide range of values of $\omega$ and $\Omega$.  Figures {\ref{fig4}}c and d illustrate the influence of the parameter $f$ on resonance for two sets of values of $\omega$ and $\Omega$.  In both the cases we observe one or more resonance for $f \ll 1$.   The values of $g$ at which resonance occur and the corresponding value of $Q$  vary with the parameters of the system.   

%
\section{Vibrational resonance in the two species model}
In the previous section, we have described the vibrational resonance associated with the vegetation dynamics of a single species interacting with the water table. The present section is devoted to a two phreatophyte species $A$ and $B$ interacting with the water table.
\subsection{Description of the model}  
In the two species (denoted by $A$ and $B$) model proposed by (Ridolfi et al., 2006, 2007) the species $A$ is assumed to be dominant over the species $B$. That is, in the absence of interactions with the water table, $A$ tends to its maximum density while $B$ tends to disappear. The values of $A(t)$ and $B(t)$ are normalized with respect to their maximum value. Frequently, the logistic law is chosen for the growth of $A$ and $B$. The model system is thus
\begin{eqnarray}
 \label{eq8}
   \frac{ {\mathrm{d}} A }{ {\mathrm{d}} t }
       & = & \alpha_A A \left( V_{ {\mathrm{c}} A} - A \right) , \\
\label{eq9}
   \frac{ {\mathrm{d}} B }{ {\mathrm{d}} t }
       & = & \alpha_B B \left( V_{ {\mathrm{c}} B} - A - B \right) ,
\end{eqnarray}
where $\alpha_A$ and $\alpha_B$ are the coefficients determining the response rate of $A$ and $B$ respectively and $V_{{\mathrm{c}}A}$ and $V_{{\mathrm{c}}B}$ are the carrying capacities of the species $A$ and $B$ respectively. $V_{ {\mathrm{c}} A }$ and $V_{{\mathrm{c}}B}$ depend on the depth $d(t)$ of the water table. Introducing the change of variables $\lambda=\alpha_B /\alpha_A$ and $t=\alpha_A t'$ (which makes time a dimensionless quantity), dropping the prime in $t'$ and assuming that the dynamics of $B$ depends on $A(t-\tau)$ where $\tau$ is a time-delay, the Eqs.~({\ref{eq8}}-{\ref{eq9}}) become
\begin{eqnarray}
 \label{eq10}
  \frac{ {\mathrm{d}} A}{ {\mathrm{d}} t}
       & = & A \left( V_{{\mathrm{c}} A} - A \right),\\ 
\label{eq11}
  \frac{ {\mathrm{d}} B}{ {\mathrm{d}} t}
       & = & \lambda B \left( V_{{\mathrm{c}} B } 
              - A(t-\tau) - B \right) .
\end{eqnarray}

In Eqs.~({\ref{eq10}}-{\ref{eq11}})
\begin{equation}
 \label{eq12}
     V_{ {\mathrm{c}} i}
     = \theta  \left[ d - d_{ {\mathrm{min}},i } \right]
          \cdot \theta \left[ d_{ {\mathrm{max}},i }
             - d \right] ,
\end{equation}
where $i=A,B$, $\theta\left[ s \right]$ is the Heaviside function, that is, $\theta \left[ s \right] =1$ for $s>0$ and $0$ for $s<0$, $d_{ {\mathrm{min}},i }$ and $d_{ {\mathrm{max}},i}$ are the minimum and maximum water table depths tolerated respectively by the species $i$ with $d_{ {\mathrm{min}},B} < d_{ {\mathrm{min}},A} < d_{ {\mathrm{max}},B }$ (species $A$ needs a deeper aquifer than species $B$) and
\begin{equation}
 \label{eq13} 
      d(t) = d_0 + \beta_A  A + \beta_B  B 
               + f \cos \omega t + g \cos \Omega t ,
\end{equation}
where $d_0$ is the water table depth in the absence of vegetation ($A=B=0$) and $\beta_A$ and $\beta_B$ are coefficients that weight the control exerted by the vegetation on the water table depth $d$.
 
In the above two species model one  of the species is assumed to be subdominant.  Further, the feedback of $A$ on the evolution of $B$ is taken as a time-delayed  term.  For a variety of phreatophytes  these assumptions can be realized.  Dominant species $A$ can be regarded as plants with tap-roots able to penetrate through relatively deeper than those of the species $B$.  In the literature of plants, approximate maximum lengths of tap-roots of phreatophytes are reported.  Some of the plants with deep penetrating  tap-roots with length more than $15{\mathrm{m}}$ are mequite, camelthorns, grease wood, and purple medic.  Plants such as black grease wood and banksia ($173$ species) have tap-roots with length in the range of $5$-$10{\mathrm{m}}$. Examples of plants with tap-roots of short length about $1$-$5{\mathrm{m}}$ are saguaro, creosote bush, ocotillo brittle bush, sagebrush, alder and chamisa.  One can identify appropriate species of types $A$ and $B$.  In this connection we wish to cite that the experimental analysis carried out on holm oaks and cork oaks indicated that the  higher water status leading to more effective drought avoidance of former is due to their deeper root systems compared to the latter (David et al., 2007).  A field experiment was performed on the two phreatophytic plant species Alhagi  sparsifolia (camelthorns) and Karelinia caspia occurring around the river Oasis at the southern fringe of the Taklamakan desert (Vonlanthen et al., 2010).  Both the species occur at sites with distances  to the ground water table upto $12{\mathrm{m}}$ while only Alhagi sparsifolia occurs at distances upto $17{\mathrm{m}}$.

The motivation for introducing the time-delay term $A(t-\tau)$ in the system (\ref{eq10}-\ref{eq11}) is to take into account  the fact that the changes in the population of a species, generally, will not have immediate effect on the growth of own population and on the interacting species.  The effect will be realized after a time-lag.  The effect of time-delay has been studied in population models and vegetation dynamics (Kuang, 1993; Wang et al., 2011).  Reports on the analysis of influence of various factors on the growth of vegetation dynamics based on AVHRR (Advanced Very High Resolution Radiometer) images indicate that the time-lag can be few days to few months (Richard and Poccard, 1998; Li et al., 2002; Wang et al., 2006; Farajzadeh et al., 2011). A reasonable value of time-delay can be of the order of $1/\alpha_B$ where $\alpha_B$  is the coefficient determining the response rate of the species $B$.  

\subsection{Multiple resonance}
For our numerical study we fix $d_0=1{\mathrm{m}}$, $d_{ {\mathrm{min}},A } = 1.5 {\mathrm{m}}$, $d_{ {\mathrm{max}},A } = 2.5 {\mathrm{m}}$, $d_{ {\mathrm{min}},B } = 0.5 {\mathrm{m}}$, $d_{ {\mathrm{max}},B } = 2 {\mathrm{m}}$, $\beta_A = 0.51{\mathrm{m}}$, $\beta_B=0.9 {\mathrm{m}}$, $\alpha_A = \alpha_B = 1 {\mathrm{d}}^{-1}$, $\omega=0.5$ and $\Omega=5$. When $f=0$ and $g=0$ the system ({\ref{eq10}}-{\ref{eq11}}) has two equilibrium states. $(A^*, B^*) = (1,0)$ is stable while $( A^*,B^*) = (0,1)$ is unstable. $(A^*, B^*) = (1,0)$ remains as a stable equilibrium point for $f<0.01$ when $\omega=0.5$, $g=0$ and $\tau=1$. The system exhibits excitable dynamics for $f>0.01$. 

We choose the value of $f$ less than $0.01$ so that in the absence of high-frequency oscillation of water table the system is in the  stable equilibrium state.  It is important to investigate the response of the system for a wide range of values of the control parameters $f$, $\omega$, $g$, $\Omega$ and $\tau$.  Because such a study  in the five parameters space is time consuming we restrict to $(g,\tau)$ parameters space.  We studied the response of the system for a wide range of values of $g$ for several fixed values of the parameters of the system.  For $g>0.4$ there are no resonances and the response amplitudes of both the species $A$ and $B$ are $\approx 0$.  Therefore, we consider the range of $g$ as $[0,0.4]$.  Before presenting  the results in $(g, \tau)$ parameter space first  we discuss the presence of resonance  for fixed values of $f$, $\omega$, $\Omega$  and $\tau$ thereby varying the parameter $g$.

Figure {\ref{fig5}} presents the numerically computed response amplitudes $Q_A (\omega)$ and $Q_B (\omega)$ as a function of $g$ for three fixed values of $f$ with $\tau=1$. There are few interesting results:
\begin{figure}[!b]     
\begin{center}
\includegraphics[width=0.5\linewidth]{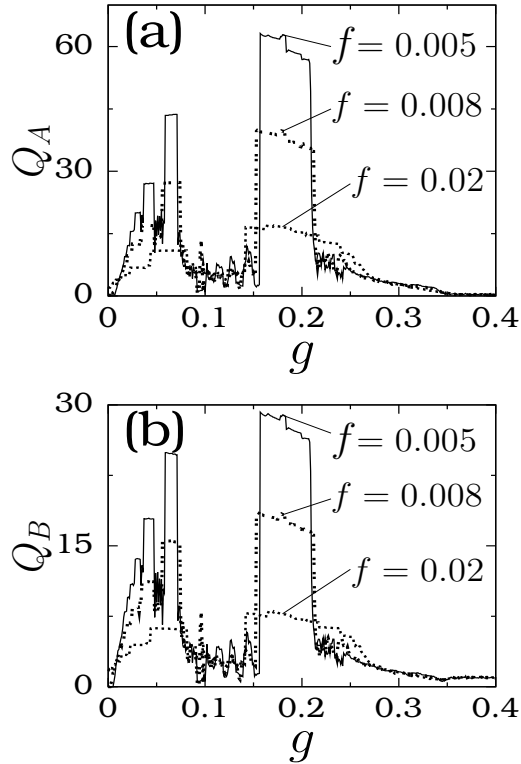}
\end{center}
\caption{Response amplitudes $Q_A$ and $Q_B$ of species $A$ and $B$ versus the parameter $g$ for three fixed values of $f$. The values of the other parameters are $d_0=1{\mathrm{m}}$, $d_{ {\mathrm{min}},A } = 1.5 {\mathrm{m}}$, $d_{ {\mathrm{max}},A } = 2.5 {\mathrm{m}}$, $d_{ {\mathrm{min}},B } = 0.5{\mathrm{m}}$, $d_{ {\mathrm{max}},B } = 2 {\mathrm{m}}$, $\beta_A = 0.51{\mathrm{m}}$, $\beta_B=0.9 {\mathrm{m}}$, $\alpha_A = \alpha_B = 1 {\mathrm{d}}^{-1}$, $\omega=0.5$, $\Omega=5$ and $\tau=1$.}
\label{fig5}
\end{figure}

\begin{itemize}
\item
The response amplitudes of the species $A$ and $B$ display similar variation, however, $Q_A > Q_B$, that is, the species $B$ is subdominant.
\item
An interesting result is that even though species $A$ is dominant, when $A$ exhibits resonance the species $B$ does not disappear but also displays a resonance and, moreover, in Fig.~{\ref{fig5}} we notice that $Q_B \approx Q_A/2$ in the resonance region. This is due to the vegetation-water table interaction.
\item
Multiple resonance occurs. $Q_A$ and $Q_B$ are not maximum at discrete values of $g$, but they are almost constant over a range of values of $g$.
\item
$Q_A$ (and $Q_B$) values at successive resonances are not equal.  In the single species model also response amplitude values are not same at successive resonances.  We wish to point out that in bistable systems driven additively by a biharmonic  force the response amplitude at successive resonances are found to be the same (Landa and McClintock, 2000; Rajasekar et al., 2010). 
\item
$Q_A$ (and $Q_B$) values at resonances decrease by increasing  the value of $f$, while the width of the resonance interval increases.
\end{itemize}

To understand the occurrence of a multiple resonance and almost plateau regions of resonance profile, we consider the nature of the evolution of the system. The system (\ref{eq10}-\ref{eq11}) is a system of two-coupled first-order nonlinear differential equations driven by a periodic force.  Such a system is capable of exhibiting different types of nonlinear dynamics including chaotic dynamics (a nonperiodic and bounded evolution of a system with high sensitive dependence on initial conditions).  In the system (\ref{eq10}-\ref{eq11}) in the absence of time-delay and high-frequency oscillation of water table chaotic dynamics is found to occur when the amplitude of the low-frequency oscillation of the water table is varied (Ridolfi et al., 2007).  However, in the system (\ref{eq10}-\ref{eq11}) for the parametric choices considered in the present work any route to  chaotic dynamics is  not observed.  

\begin{figure}[t]     
\begin{center}
\includegraphics[width=0.5\linewidth]{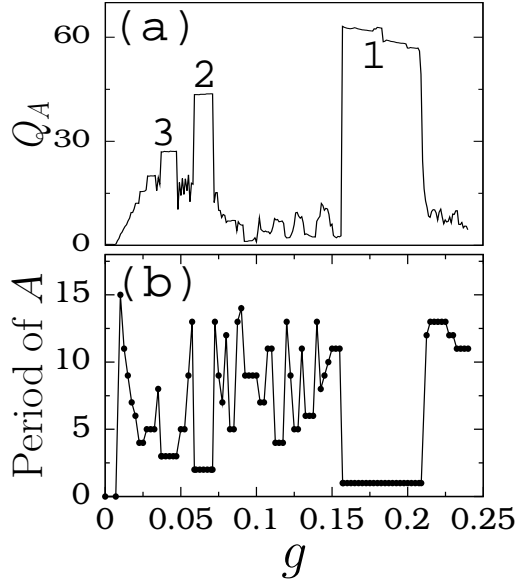}
\end{center}
\caption{Plots of (a) the response amplitude $Q_A$ and (b) the period of $A$  (in units of $T$) versus $g$ for $f=0.005$.  In the subplot (a) three most dominant intervals of $g$ with large values of $Q_A$ are marked by the numbers $1$, $2$ and $3$. The values of the other parameters in the system  (\ref{eq10}-\ref{eq11}) are as in Fig.~\ref{fig5}. }
\label{fig6}
\end{figure}

In Fig.~\ref{fig5} the variation of $Q_A$ and $Q_B$ is shown for $g \in [0,\,0.4]$ only.  For $g>0.4$, as mentioned earlier, we found $Q_A$ and $Q_B \approx 0$ and hence we restrict ourselves to the interval $0<g<0.4$.  In this interval of $g$  either stable equilibrium state or periodic variation of $A$ and $B$ is found depending upon the value of $g$. To identify the period of $A(t)$ we collect the values of $A(t)$ at $t=n T$ where $n=1,2,\cdots,m(=500)$, $T=2 \pi / \omega$ and designate them as $A_n$.  If $A(t)$ is periodic with period $T$ then $A_1=A_2 = \cdots =A_m$.  For a period-$2T$ variation of $A(t)$ we observe  $A_1=A_3=\cdots=A_{m-1}=a_1$, $A_2=A_4=\cdots=A_{m}=a_2$ and   $a_1 \ne a_2$.  Similarly, we can define higher periods of $A$ and identify the periodicity of $A$.   In Fig.~{\ref{fig6}} we have plotted the period of $A$ together with $Q_A$ for $f=0.005$. The period of $A$ is a complicated function of $g$. 

\begin{figure}[t]     
\begin{center}
\includegraphics[width=0.82\linewidth]{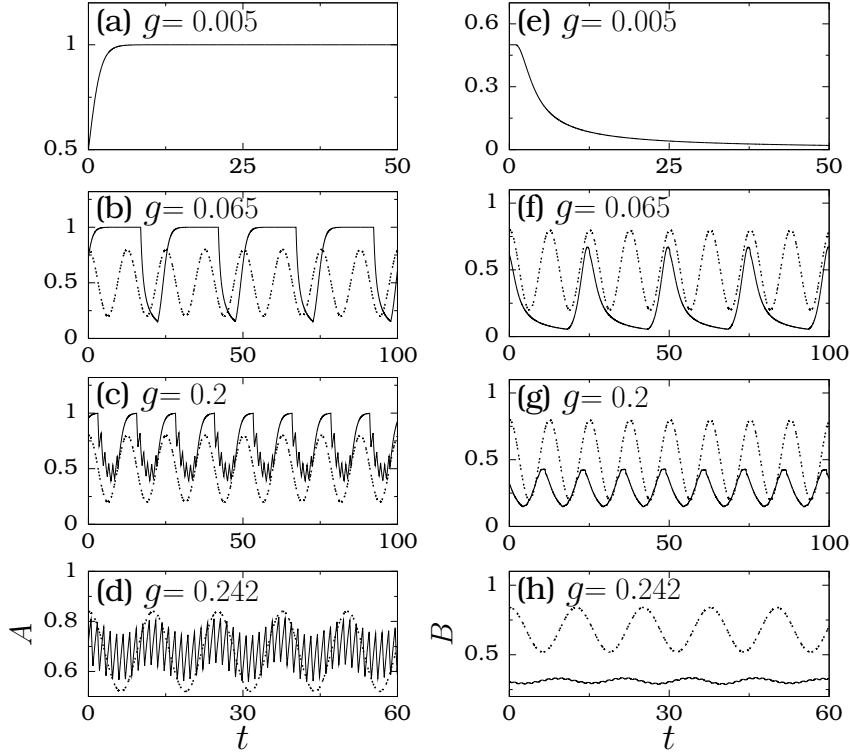}
\end{center}
\caption{Temporal behaviour of (a)-(d) $A$  and (e)-(h) $B$ for four fixed values of $g$ with $f=0.005$.  The values of the other parameters in the system (\ref{eq10}-\ref{eq11}) are as in Fig.~\ref{fig5}. The dashed curve represents $f \cos \omega t$ and its amplitude is suitably magnified and its center of oscillation is shifted for illustrative purpose.}
\label{fig7}
\end{figure}

Interestingly, the period remains the same in the regions of $g$ where $Q$ is almost a constant. Different plateau regions correspond to different constant  periods of $A$. For $g < 0.007$ we observe $(A,B) \to (1,0)$ as $t$ increases. The species biomass $A$ attains its maximum while that of the species $B$ failed to survive. Even though the water table depth $d(t)$ varies periodically, the evolution of $A$ and $B$ is not oscillatory but approaches the equilibrium state for $g < 0.007$ as shown in Figs.~{\ref{fig7}}a and e for $g=0.005$.
This is because the oscillatory variation of $d(t)$ does not strictly appear as an additive driving force in ({\ref{eq10}}-{\ref{eq11}}) but its influence is on $V_{ {\mathrm{c}} A }$ and $V_{ {\mathrm{c}} B }$. Equations ({\ref{eq10}}-{\ref{eq11}}) can be regarded as a parametrically driven system. Moreover, $V_{ {\mathrm{c}} A }$ and $V_{ {\mathrm{c}} B }$ are not sinusoidally varying function of time. They take the values $0$ or $1$ depending upon the values of the various parameters of the system and the value of $t$.

The most dominant resonance interval is $g \in [0.157,0.21]$. In this interval of $g$ the period of $A$ is $T$ (see Fig.~{\ref{fig6}}b). Figures~{\ref{fig7}}c and g show the evolution of $A$ and $B$ with time for $g=0.2$. We can clearly notice the best synchronization of $A$ and $B$ with the external drive  $f \cos \omega t$. The next dominant resonance interval is $g \in [0.059,0.071]$. Figures {\ref{fig6}}b indicates that the period of $A$ in this interval of $g$ is $2T$. In Figs.~{\ref{fig7}}b and {\ref{fig7}}f for $g=0.065$ we observe pulse-like solution with period-$2T$. Period-$3T$ oscillation of $A$ and $B$ occurs in the interval $g \in [0.037,0.047]$ (not shown in Fig.~\ref{fig7}) and this region constitutes the third dominant resonance region.

For $g>0.24$ the evolution of $A$ and $B$ exhibits a rapid oscillation (as shown in Figs.~{\ref{fig7}}d and h for $g=0.242$) and both $Q_A$ and $Q_B$ decays to zero implying degrading of the response of the biomasses $A$ and $B$ with increase in the value of $g$. Outside the three dominant resonance regions, we observe few small intervals of $g$ where $Q$ becomes almost constant but relatively very small due to higher periodicity of variation of $A$ and $B$. In Fig.~{\ref{fig7}} we notice that as $g$ increases from a small value the evolution of $A$ and $B$ undergoes a transition from equilibrium state $\to$ pulse-like solution $\to$ rapid oscillatory solution. Further, when $A$ becomes maximum (minimum) $B$ becomes minimum (maximum).  From the above, we point out the following features:
\begin{itemize}
\item
$Q$ is very small if the species biomass remains a constant (see Figs.~\ref{fig7}a and e) or periodic with large period or it oscillates rapidly (see Figs.~\ref{fig7}d and h).
\item
$Q$ is larger, for example the regions $1$, $2$ and $3$ in Fig.~\ref{fig6}a, when the  period of the pulse-like  evolution of the  biomass species is lower.
\item
$Q$ remains almost constant over an interval of $g$ (regions $1$, $2$ and $3$ of Fig.~\ref{fig6}a) if the period of $A$ remains the same.
\end{itemize}

It is noteworthy to compare the mechanisms of vibrational resonance in the FitzHugh--Nagumo (FN) equation, a well studied excitable system, and the system ({\ref{eq10}}-{\ref{eq11}}). In the FN equation when the amplitude of the high-frequency force varies, a resonance takes place when the waiting time $(T_{{\mathrm{w}}})$, the time the system spends around the equilibrium point between two consecutive firing, is $T/2$ (Ullner et al., 2003; Hu et al., 2012). In the noise-induced stochastic resonance also a resonance occurs when $T_{{\mathrm{w}}} \approx T/2$. In the system ({\ref{eq10}}-{\ref{eq11}}), $Q$ is inversely proportional to the period of evolution of biomasses.

\subsection{Effect of a time-delay on the resonance}
So far, we have focused our analysis for a specific value of the time-delay $\tau(=1)$. Now, we present the effect of time-delay on the resonance.

\begin{figure}[!h]     
\begin{center}
\includegraphics[width=0.45\linewidth]{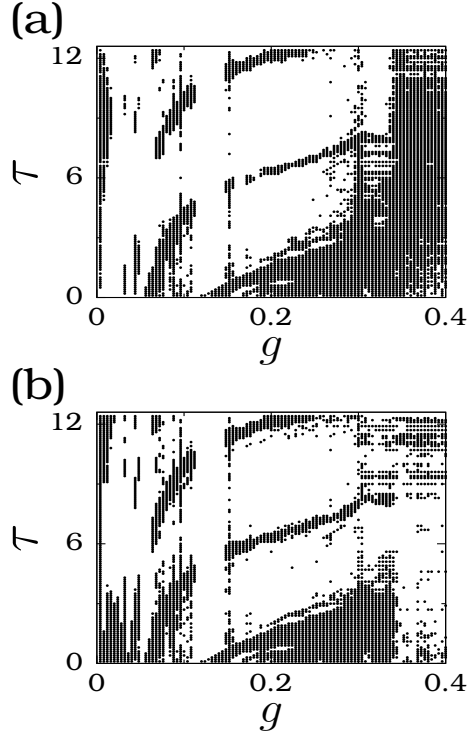}
\end{center}
\caption{Regions (marked by black colour) where (a) $Q_A(\tau,g) > Q_A(\tau=0,g)$ and (b) $Q_B(\tau,g) > Q_B(\tau=0,g)$ for the system ({\ref{eq10}}-{\ref{eq11}}).  The values of the parameters are as in Fig.~\ref{fig5}. } 
\label{fig8}
\end{figure}

First, in the parameter space $(g,\tau)$, we identify the regions where $Q(\tau,g) > Q(0,g)$ for both species. We divide the $(g,\tau)$ parameter space in the interval $g \in [0,0.4]$ and $\tau \in [0,12]$ into $100 \times 100$ grid points. We collect the grid points for which $Q(\tau,g) > Q(0,g)$. The result is the Fig.~{\ref{fig8}}.  Both Figs.~{\ref{fig8}}a and b corresponding to the species $A$ and $B$ respectively are almost similar except near $g=0$ (far before resonance) and near $g=0.4$ (far after the last resonance) and in these two regions $Q_A$ and $Q_B$ are very small. The time-delay $\tau$ has a strong influence on the response amplitudes of $A$ and $B$.

\begin{figure}[t]     
\begin{center}
\includegraphics[width=0.5\linewidth]{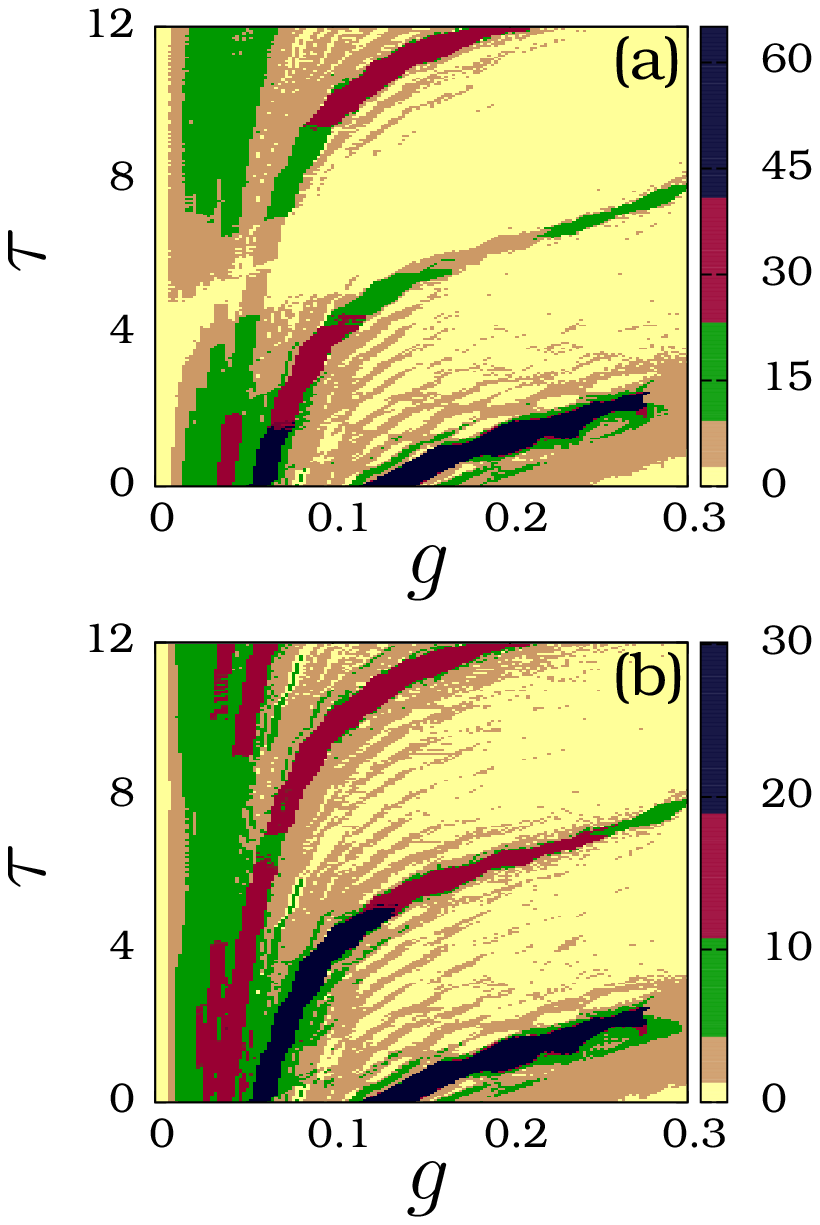}
\end{center}
\caption{(Colour in online) Colour-coded dependence of $Q_A$ and $Q_B$ of the system ({\ref{eq10}}-{\ref{eq11}}) on the parameters $\tau$ and $g$ for $f=0.005$. The values of other parameters are as in Fig.~\ref{fig5}. }
\label{fig9}
\end{figure}

\begin{figure}[t]     
\begin{center}
\includegraphics[width=0.65\linewidth]{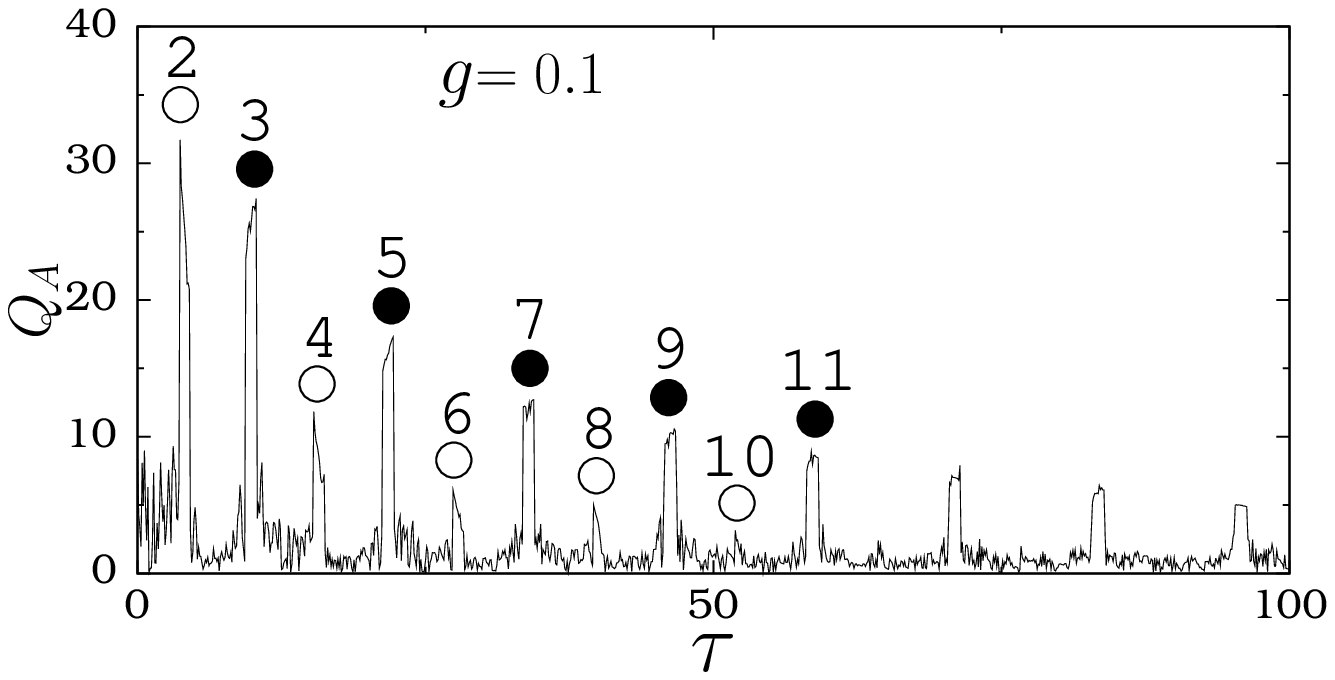}
\end{center}
\caption{$Q_A$ of the system ({\ref{eq10}}-{\ref{eq11}}) as a function of time-delay $\tau$ for $g=0.1$.  The values of other parameters are as in Fig.~\ref{fig5}.  The period of $A$ in units of $T(=2 \pi / \omega)$ is marked for first few resonance intervals.  The resonance intervals with period of $A$ equal to even (odd) integer multiples of $T$ are marked by open (solid) circles.  }
\label{fig10}
\end{figure}

Next, Fig.~{\ref{fig9}} features the colour-contour plots of the dependence of  $Q_A$ and $Q_B$ on $g$ and $\tau$.  Multiple resonance is found to occur for a wide range of fixed values of $\tau$. Moreover, Fig.~{\ref{fig9}} clearly demonstrates the strong influence of $\tau$ on the variation of $Q$. 
For fixed values of $g$ the response amplitudes $Q_A$ and $Q_B$ are not periodic with $\tau$. Figure {\ref{fig10}} depicts the dependence of $Q_A$ on the time-delay $\tau$ for $g=0.1$.  $Q_A$ (as well as $Q_B$) exhibits a sequence of resonance intervals.  The period of $A$ (as well as $B$) in the first resonance interval is $2T$. 
The periods of $A$ in the other consecutive intervals are $3T$, $4T$, $\cdots$ respectively.  That is, the system exhibits a resonance sequence with period adding dynamics.  The resonance intervals with period of $A$ being even (odd) integer multiples of $T$ are marked by open (solid) circles in Fig.~\ref{fig10}.  The values of $Q_A$ of these two sequence of resonance intervals decrease rapidly with $\tau$.  Further, we observe that $Q_A(\tau)$ does not varies periodically with $\tau$, that is $Q_A(\tau+\alpha) \ne Q_A(\tau)$ for some finite nonzero value of $\alpha$. However, the resonance intervals marked by solid circles and open circles occur at a regular interval of delay-time $\approx T$.  Similar dependence of $Q$ on $\tau$ is found to occur for other fixed values of $g$. We note that in nonlinear oscillators with time-delayed feedback and driven additively by a biharmonic force $Q$ is shown to vary periodically (Yang and  Liu, 2010;  Jeevarathinam et al., 2011).

In this section we presented our analysis on the system (\ref{eq10}-\ref{eq11}) for $\omega=0.5$ and $\Omega=10$.  Results similar to these fixed values of $\omega$ and $\Omega$ are observed  for a wide range of values of $\omega$ and $\Omega$.  The number of resonances, the values of $Q$ at resonances and the values of $g$ at which resonances occur depend on the values of the other parameters of the system.

\section{Conclusion}
Several studies have reported the occurrence of vibrational resonance induced by a two-frequency periodic force in physical and biological nonlinear systems. The present work is devoted to the analysis of the effect of a biharmonic type variation of the water table depth in two phreatophyte plant ecosystems. We have considered a single species and a two species model systems. The former has bistability, while the later is an excitable system. Our study shows how a very simple deterministic periodic variation of the water depth is able to give rise to  a great increase in the response of vegetation dynamics of single species and two species ecosystems.

In both overdamped and underdamped nonlinear oscillators exhibiting multiple vibrational resonance (Landa and McClintock, 2000; Baltanas et al., 2003; Jeyakumari et al., 2009; Yang and Liu, 2010; Jeevarathinam et al., 2011), at all resonances the long time motion of the systems is found to be periodic with period $T$ of the low-frequency force when the ratio $\Omega / \omega$ is an integer. In the single species system (\ref{eq1}) the period of evolution is $T$ at all the resonances, however, the value of $Q$ at successive resonances decreases. In the two species model, the period of the system is different at resonances. Further, the response amplitude $Q$ is found to be inversely proportional to the period of the evolution of the species biomass. Analysis of vibrational resonance in systems described by different kinds of evolution equations can lead to a deeper understanding of the phenomenon.

We believe that analysis of nonlinear phenomena such as chaos, stochastic and vibrational resonances, in the theoretical models of vegetation-water table interactions may motivate the experimentalists to perform experiments with controlled variations of the water table depth in a small scale. Such studies not only would explore the ways of enhancing the biomass response, but also help to improve the theoretical models based on experimental observations.  

\subsection*{{\bf{Acknowledgments}}}
CJ acknowledges the support from University Grants Commission, India  in the form of UGC-Meritorious Fellowship. One of us (SR) would like to thank   M.~Sundararaman for his helpful discussions. Financial support from the Spanish Ministry of Science and Innovation under Project No. FIS2009-09898 is acknowledged by MAFS.  We gratefully thank anonymous referees for their constructive comments and suggestions on  our earlier draft which improved the quality and presentation of the paper.

\section*{References}    
\begin{description}
\item{}
Baltanas, J.P., Lopez, L., Blechman, I.I., Landa, P.S., Zaikin, A., Kurths, J., Sanjuan, M.A.F., 2003. Experimental evidence, numerics, and theory of vibrational resonance in bistable systems. Phys. Rev. E 67, 066119-1-7. 
\item{}
Blekhman, I.I., Landa, P.S., 2004. Conjugate resonances and bifurcations in nonlinear systems under biharmonical excitation. Int. J. Nonlinear Mech. 39, 421-426.
\item{}
Borgogno, F., D'Odorico, P., Laio, F., Ridolfi, L., 2012. Stochastic resonance and coherence resonance in groundwater-dependent plant ecosystems. J. Theor. Biol. 293, 65-73.
\item{}
Chizhevsky, V.N., Giacomelli, G., 2006. Experimental and theoretical study of vibrational resonance in a bistable system with asymmetry. Phys. Rev. E 73, 022103-1-4.
\item{}
David, T.S., Henriques, M.O., Kurz-Besson, C., Nunes, J., Valente, F., Vaz, M., Pereira, J.S., Siegwolf, R., Chaves, M.M., Gazarini, L.C., David, J.S., 2007. Water-use strategies in two-co-occurring Mediterranean evergreen Oaks: surviving the summer drought. Tree Physiology 27, 793-803.
\item{}
D'Odorico, P., Laio, F., Ridolfi, L., Lerdau, M.T., 2008. Biodiversity enhancement induced by environmental noise. J. Theor. Biol. 255, 332-337. 
\item{}
Elmore, A.J., Manning, S., Mustard, J.F., Craine, J.M., 2006. Decline in alkali meadow vegetation cover in California: the effects of groundwater extraction and drought. J. Appl. Ecol. 43, 770-779.
\item{}
Farajzadeh, M., Fathnia, A., Alijani, B., Zeaiean, P., 2011. Assessment of the effect of climatic factors on the growth of dense pastures of Iran using AVHRR images. Phys. Geog. Res. Quarterly 75, 1-2. 
\item{}
Gammaitoni, L., Hanggi, P., Jung, P., Marchesoni, F., 1998. Stochastic resonance. Rev. Mod. Phys. 70, 223-287.
\item{}
Gandhimathi, V.M., Rajasekar, S., Kurths, J., 2006. Vibrational and stochastic resonances in two coupled overdamped anharmonic oscillators. Phys. Lett. A 360, 279-286.
\item{}
Hu, D., Yang, J.H., Liu, X.B., 2012. Delay-induced vibrational multiresonance in FitzHugh-Nagumo system. Commun. Nonlinear Sci. Numer. Simulat. 17, 1031-1035.
\item{}
Jeevarathinam, C., Rajasekar, S., Sanjuan, M.A.F., 2011. Theory and numerics of vibrational resonance in Duffing oscillators with time-delayed feedback. Phys. Rev. E 83, 066205-1-12.
\item{}
Jeyakumari, S., Chinnathambi, V., Rajasekar, S., Sanjuan, M.A.F., 2009. Single and multiple vibrational resonance in a quintic oscillator with monostable potentials. Phys. Rev. E 80, 046608-1-8.
\item{}
Kuang, Y., 1993. Delay Differential Equations with Applications in Population Dynamics. Academic Press, Boston.
\item{}
Lai, Y.C., Liu, Y.R., 2005. Noise promotes species diversity in nature. Phys. Rev. Lett. 94, 038102-1-4.
\item{}
Landa, P.S., McClintock, P.V.E., 2000. Vibrational resonance. J. Phys. A: Math. Gen. 33, L433-L438.
\item{}
Li, L., Barry, D.A., Parlange, J.Y., Pattiaratchi, C.B., 1997. Beach water table fluctuations due to wave runup: Capillarity effects. Water Res. Res. 33, 935-945.
\item{}
Li, B., Tao, S., Dawson, R.W., 2002. Relation between AVHRR NDVI and ecoclimatic parameters in China. Int. J. Remote Sensing 23, 989-999. 
\item{}
McDonnell, M.D., Stocks, N.G., Pearce, C.E.M., Abbott, D., 2008. Stochastic Resonance: From Suprathreshold Stochastic resonance to Stochastic signal Quantisation, Cambridge University Press, Cambridge.
\item{}
Munoz-Reinoso, J.C., de Castro, F.J., 2005. Application of a statistical water-table model reveals connections between dunes and vegetation at Donana.  J. Arid. Environ. 60, 663-679.
\item{}
Naumburg, E., Mata-Gonzalez, R., Hunter, R.G., Mclendon, T., Martin, D.W., 2005. Phreatophytic vegetation and groundwater fluctuations: a review of current research and application of ecosystem response modeling with an emphasis on great basin vegetation. Environ. Manage. 35, 726-740.
\item{}
Pikovsky, A.S., Kurths, J., 1997. Coherence resonance in a noise-driven excitable system. Phys. Rev. Lett. 78, 775-778.
\item{}
Rajasekar, S.R., Jeyakumari, S., Chinnathambi, V., Sanjuan, M.A.F., 2010. Role of depth and location of minima of a double-well potential on vibrational resonance.  J. Phys. A: Math. Theor. 43, 465101-1-11.
\item{}
Richard, Y., Poccard, I., 1998. A statistical study of NDVI sensitivity to seasonal and interannual rainfall variations in southern Africa. Int. J. Remote Sensing 19, 2907-2920. 
\item{}
Ridolfi, L., D'Odorico, P., Laio, F., 2006. Effect of vegetation-water table feedbacks on the stability
and resilience of plant ecosystems. Water Resour. Res. 42, W01201-W01205.
\item{}
Ridolfi, L., D'Odorico, P., Laio, F., 2007. Vegetation dynamics induced by phreatophyte-aquifer interactions. J. Theor. Biol. 248, 301-310.
\item{}
Roxburgh, S.H., Shea, K., Wilson, J.B., 2004. The intermediate disturbance hypothesis: patch dynamics and mechanics of species coexistence.
Ecology 85, 359-371.
\item{}
Scarsoglio, S., D'Odorico, P., Laio, F., Ridolfi, L., 2012. Spatio-temporal stochastic resonance induces patters in wetland vegetation dynamics. Ecological Complexity 10, 93-101.
\item{}
Scheffer, M., Carpenter, S., Foley, J.A., Folke, C., Walker, B., 2001. Catastrophic shifts in ecosystems. Nature 413, 591-596.
\item{}
Steneck, R.S., Graham, M.H., Bourque, B.J., Corbett, D., Erlandson, J.M., Estes, J.A., Tegner, M.J., 2002. Kelp forest ecosystems: biodiversity, stability, resilience and future. Environ. Conserv. 29, 436-459.
\item{}
Ullner, E., Zaikin, A., Garcia-Ojalvo, J., Bascones, R., Kurths, J., 2003. Vibrational resonance and vibrational propagation in excitable systems. Phys. Lett. A 312, 348-354.
\item{}
Vasseur, D.A., Fox, J.W., 2007. Environmental fluctuations can stabilize food web dynamics by increasing synchrony. Ecol. Lett. 10, 1066-1074.
\item{}
Vonlanthen, B., Zhang, X., Bruelheide, H., 2010.  On the run for water - Root growth of two phreatophytes in the Taklamakan desert. J. Arid Environments 74, 1604-1615.
\item{}
Waddell, E., 1976. Swash-groundwater-beach profile interactions. SEPM Spec. Publi. 24, 115-125.
\item{}
Wang, W., Anderson, B.T., Phillips, N., Kaufmann, R.K., Potter, C., Myneni, R.B., 2006. Feedbacks of vegetation on summertime climate variability over the North American Graselands. Part-I $-$ Statistical Analysis. Earth Interactions 10, 1-27 (paper No. 17).
\item{}
Wang, K., Zhang, N., Niu, D., 2011. Periodic oscillations in a spatially explicit model with delay effect for vegetation dynamics in freshwater marshes. J. Biol. Systems 19, 131-147.
\item{}
Yang, J.H., Liu, X.B., 2010. Delay induces quasi-periodic vibrational resonance. J. Phys. A: Math. Theor. 43, 122001-1-7.
\item{}
Yao, C., Zhan, M., 2010. Signal transmission by vibrational resonance in one-way coupled bistable systems. Phys. Rev. E 81, 061129-1-8.
\item{}
Yu, H., Wang, J., Liu, C., Deng, B., Wei, X., 2011. Vibrational resonance in excitable neuronal systems. Chaos 21, 043101-1-10.
\item{}
Zaikin, A.A., Lopez, L., Baltanas, J.P., Kurths, J., Sanjuan, M.A.F., 2002. Vibrational resonance in a noise-induced structure. Phys. Rev. E 66, 011106-1-9.
\end{description}

\end{document}